%% file: 00_main.tex
\pdfoutput=1

\documentclass{vgtc}                          

\ifpdf
  \pdfoutput=1\relax                   
  \usepackage{graphicx}                
  \DeclareGraphicsExtensions{.pdf,.png,.jpg,.jpeg} 
\else
  \usepackage{graphicx}                
  \DeclareGraphicsExtensions{.eps}     
\fi%

\graphicspath{{figures/}{pictures/}{images/}{./}} 

\usepackage{microtype}                 
\PassOptionsToPackage{warn}{textcomp}  
\usepackage{textcomp}                  
\usepackage{mathptmx}                  
\usepackage{times}                     
\usepackage{cite}                      
\usepackage{tabu}                      
\usepackage{booktabs}                  

\usepackage{tabularx}
\usepackage{xcolor}
\usepackage{colortbl}
\usepackage[utf8]{inputenc}
\usepackage{verbatim} 
\usepackage{xspace}

\usepackage{todonotes}

\usepackage{amssymb}

\usepackage[inline]{enumitem}
\setlist[enumerate,1]{label=\textit{\alph*)}}

\usepackage{inconsolata}
\usepackage{balance}
\usepackage{dblfloatfix}

\usepackage{scalerel}

\onlineid{0}

\vgtccategory{Research}

\vgtcinsertpkg

\title{\app: A Scalable Constructive Visualization Tool}

 \author{Gonzalo Gabriel M\'{e}ndez\thanks{e-mail: gmendez@espol.edu.ec}\\ %
 	\parbox{2in}{\scriptsize \centering Escuela Superior Polit\'{e}cnica del Litoral\\University of Calgary} %
 \and Jagoda Walny\thanks{e-mail: jkwalny@ucalgary.ca}\\ %
      \scriptsize University of Calgary%
 \and S{\o}ren Knudsen\thanks{e-mail: sknudsen@ucalgary.ca}\\ %
      \scriptsize University of Calgary%
  \and Charles Perin\thanks{e-mail: cperin@uvic.ca}\\ %
  \scriptsize University of Victoria\\[0.2in]%
  \and Samuel Huron\thanks{e-mail: samuel.huron@telecom-paristech.fr}\\ %
  \scriptsize T\'{e}l\'{e}com Paristech, Universit\'{e} Paris-Saclay%
  \and Jo Vermeulen\thanks{e-mail: jo.vermeulen@cs.au.dk}\\ %
  \scriptsize Aarhus University%
  \and Richard Pusch\thanks{e-mail: rapusch@ucalgary.ca}\\ %
  \scriptsize University of Calgary%
  \and Sheelagh Carpendale\thanks{e-mail: sheelagh@sfu.ca}\\ %
  \scriptsize Simon Fraser University%
}

\abstract{Constructive approaches to visualization authoring have been shown to offer advantages such as providing options for flexible outputs, scaffolding and ideation of new data mappings, personalized exploration of data, as well as supporting data understanding and literacy. However, visualization authoring tools based on a constructive approach do not scale well to larger datasets. As construction often involves manipulating small pieces of data and visuals, it requires a significant amount of time, effort, and repetitive steps. We present \app, an authoring tool in which a visualization is constructed by instantiating its structural and functional components through four interaction elements (objects, modifiers, activators, and tools). This design offers a new balance between preserving the benefits of a constructive process and incorporating a new approach to scalability issues. It allows designers to propagate individual mapping steps to all the elements of a visualization. 
}

\CCScatlist{
  \CCScatTwelve{Human-centered computing}{Visualization}{Visualization systems and tools}{};
  \CCScatTwelve{Human-centered computing}{Human computer interaction (HCI)}{Interactive systems and tools}{}
}

\usepackage{subcaption}

\definecolor{brightmaroon}{rgb}{0.76, 0.13, 0.28}

\newcommand{\app}{ReConstructor}
\newcommand*\quotes[1]{``#1''}

\newcommand{\prop}[1]{{\fontsize{8.5}{10}\fontfamily{zi4}\selectfont#1}}
\newcommand{\propCaption}[1]{{\fontsize{7}{10}\fontfamily{zi4}\selectfont#1}}

\setlist[itemize]{leftmargin=*,labelsep=0.25em,itemsep=0em, topsep=0em,parsep=0.6em,label=\tiny$\blacksquare$}%

\begin{document}


\maketitle

%


\input{sections/01.Introduction}

\input{sections/02.Background.tex}

\input{sections/03.Deconstruction.tex}
\input{sections/05.SystemDescription.tex}
\input{sections/09.Discussion.tex}

\input{sections/10.Conclusion.tex}
\input{sections/11.Acknowledgements.tex}


\bibliographystyle{abbrv-doi}

\bibliography{bibliography}
\end{document}

%% file: sections/01.Introduction.tex
\section{Introduction}

Visualization authoring tools give people varying degrees of control over how their data is visually represented even when they do not have enough time, resources, or skills to make custom visualizations programmatically. There are a growing number of these tools and, correspondingly, a growing number of approaches to designing them.

\noindent Popular tools, such as Excel~\cite{MSExcel} and Tableau Desktop~\cite{Tableau}, are primarily based on chart templates, automated mappings, and recommendation systems. These features are key for achieving a speedy authoring process and are beneficial for accessibility and ease of use. However, they can also impose barriers. Even for those with formal training in visualization, some types of automated approaches (such as templates) can be a hurdle to flexibility, get in the way of ideation processes, and may even take over the design lead. Recent research suggests that a constructive approach to visualization~\cite{Huron.2014a} has the potential to avoid some of these problems.\\

\noindent Constructive visualization promotes the idea of creating visual representations of data by assembling building blocks that are mapped to specific aspects of a dataset. This authoring strategy has been shown to empower people in their use of visualization without the need of formal training or other specialized skills (e.g., programming). Studies suggest that digital constructive visualization tools  can also promote a mindful design process in which people are encouraged to actively reflect on their design decisions~\cite{Mendez2017, Mendez2018}. \\

\noindent Despite these advantages, the authoring process of existing constructive tools---both tangible~\cite{huron2016, huron.2014b} and digital~\cite{Mendez2016}---does not scale well to larger datasets. Even for moderately simple visualizations, construction requires a significant amount of time, effort, and repetitive interactions~\cite{Mendez2017, Wun2016}. The problem is exacerbated as the amount of data to be represented (i.e., total records and attributes) grows. 


\noindent Conventional visualization tools avoid the scalability issues of their constructive counterparts by automating certain steps of the visualization process. While automation may reduce the effort required from the designer, it tends to also interfere with personal agency. Automation can make the authoring process less incremental as the tasks delegated to the tool are often executed quickly and over large portions of data (e.g., entire attributes rather than individual values). This, in turn, reduces the transparency of the design process, as the tool's actions can be hard to follow, which can interfere with comprehension.\\

\noindent We join the growing body of work that investigates how to expand visualization authoring options. In particular, we present \app, a new point of exploration within the design space of digital constructive visualization tools. In \app, visualizations are constructed by instantiating their structural and functional components through the use of four interaction elements (\textit{objects}, \textit{modifiers}, \textit{activators}, and \textit{tools}). This design allows people create visualizations via a user-driven constructive approach that also eases the difficulty of working with larger datasets. That is, \app~supports a more scalable authoring process while keeping the agency of this process on the user's side.

\noindent More specifically, our work contributes: (1) a construction strategy for the design of scalable visualization tools based on four reusable interaction elements: objects, modifiers, activators and tools; (2) an explanation of how these interaction elements can be incorporated in the design of visualization authoring tools; and (3) the design and implementation of \app, a tool that supports the construction of visualizations through these elements.

%% file: sections/02.Background.tex
\section{Related Work}
\label{section:relatedwork}

In this section, we discuss how the concept of construction has been used in the design of computer interfaces and, in particular, of recent visualization authoring tools.

\subsection{Constructive Theories}

Educational and learning theories such as Piaget's constructivism~\cite{Ackermann2001, Piaget2013}, Papert's constructionism ~\cite{Papert1991}, and Froebel's \textit{gifts}~\cite{stiny1980kindergarten} suggest that one way that humans discover the world is by manipulating simple objects and that we can construct knowledge and meaning from these experiences. These theories focus on personal experience, \quotes{\textit{where the learner is consciously engaged in constructing a public entity}}~\cite[p. 1]{Papert1991}, as the gateway to understanding and reflection. 

\noindent Computational tools that implement constructive principles generally support processes based on an incremental, bottom-up strategy. This is related to the concept of \textit{emergence}, by which a complex entity arises as the result of interactions among smaller or simpler entities~\cite{abbott2004emergence}. Consequently, a constructive process is beneficial to show how a complex structure is built from the ground up, as the result of many small steps or sub-processes.\\


\noindent Constructive theories have been widely explored in environments that support the development of computational thinking skills such as Scratch~\cite{Dasgupta2015, maloney2010scratch, resnick2007all, Resnick} and Mindstorms~\cite{Papert1993} and, more recently, Google's Blockly library~\cite{Google2017}. In these tools, construction takes place with building blocks that animate interactive visuals.

\subsection{Construction in InfoVis}

Constructive Visualization~\cite{Huron.2014a}, a paradigm for visualization authoring grounded in Papert's, Piaget's and Froebel's theories, imports from them the idea of using physical \textit{tokens} (e.g., Lego blocks) that can be mapped to data and manipulated to compose tangible representations. iVoLVER~\cite{Mendez2016} implements constructive principles in a digital visual programming environment. Both Huron et al.'s tangible tokens (\cite{huron2016, huron.2014b}) and iVoLVER's marks support work with atomic data elements (e.g., individual values as opposed to entire data attributes). This leads to a bottom-up construction strategy in which the final design emerges as the result of several small-scale decisions and manipulations of the visualization elements.

\noindent Constructing visual representations of data from their atomic building blocks is in line with Bertin's semiotic views~\cite{Bertinb, Bertina}. To convey messages visually, Bertin's marks---graphical primitives such as points or lines---are configured in particular ways by mapping data attributes to their visual properties~\cite{Cleveland1984, Cleveland1987}. 

\noindent \app's constructive approach also makes use of building blocks to provide access to a visualization's component (e.g., marks and visual properties) and to represent the operations that take place in different types of visual encodings (e.g., sorting, spacing).

%% file: sections/03.Deconstruction.tex
\section{Enabling Construction}

For a visualization to be \textit{constructed}, the first step is to \textit{deconstruct} it into its modular components. We take this step to make use of the incremental nature of constructive visualization~\cite{Huron.2014a}. 
To support the construction of a given visualization, the design of \app requires to identify the objects involved (e.g., visual marks, axes), their attributes (e.g., visual properties, labels), and their associated functionality (i.e., the processes associated to these objects such as \textit{sorting} or \textit{distributing}). We also have to pay attention to how these components interact. Having the visualization's objects, attributes, and functionality as modular components enables a constructive approach to visualization authoring.

\subsection{Interaction Elements}

\app's design is based on four fundamental interaction elements: objects, modifiers, activators, and tools. When combined, these elements enable reconstruction of a visualization from its modular components.\\

%
%

\noindent
An \textbf{object} 
has a graphical representation, which supports visibility~\cite{norman:1988:psychology_everyday_things, norman:2010:gestural_usability} and allows for direct manipulation~\cite{Shneiderman1983}. They have various visible attributes (e.g., \prop{fill} and \prop{stroke} colors) and are by default inert---they do not cause any effect or interact with each other.\\

\noindent A \textbf{modifier} can take action on an object, such as applying the value of the object's attribute. It has the ability to change an object.
For example, a \prop{fill} modifier can change the fill color of an object and a \prop{stroke} modifier changes the color of the stroke that a pen tool produces. Modifiers can be of two types:
\vspace{-\parskip}
\begin{itemize}[leftmargin=0.6em, labelsep=0.15em]
	\setlength\itemsep{-0.5em}
	
	\item[-] \textit{Transient} modifiers change an object's attribute when placed on the object. Because the change in the object is visually conveyed, these modifiers are not graphically represented on the canvas. Transient modifiers include the \prop{stroke} and \prop{fill} color modifiers, as well as the \prop{shape} modifier that turns strokes into regular shapes. 
	
	\item[-] \textit{Persistent} modifiers do not immediately change an attribute when placed on an object. Instead, they remain visible, attached to the object they modify. That is, they ``objectify''~\cite{Xia2016} attributes (e.g., \prop{width} or \prop{height}). Dropping a data dimension from a dataset onto the visual representation of a persistent modifier establishes a mapping between the corresponding data dimension and attribute. Persistent modifiers can add properties that are not inherent to strokes. For example, the \prop{label} modifier adds text to an object.\\
\end{itemize}

\noindent An \textbf{activator} carries a process that can be dynamically added to an object. That is, activators bring inert objects to life by turning them into \textit{tools}. For example, a \prop{push} activator carries the process of \prop{pushing} other objects around. Adding a \prop{push} activator to a squiggly stroke would give it the ability to push other objects when dragged. This activating strategy is relates to ActiveInk~\cite{Romat:2019:ActiveInk}, where ink annotations can be activated to operate on visual representations.\\ 

\noindent A \textbf{tool} is an object that has been activated. A tool can have one or more activators associated with it. For example, adding an \prop{ink} activator to a \prop{push}-activated stroke results in a pen that draws while pushing other objects out of the way.
Activators and persistent modifiers can also be removed from tools. For example, when a \prop{push} activator is removed from an object it no longer pushes other objects. Removing one activator does not affect the functionality added by other activators (if any).



\begin{figure}[b!]
	\begin{center}
		\includegraphics[width=\columnwidth]{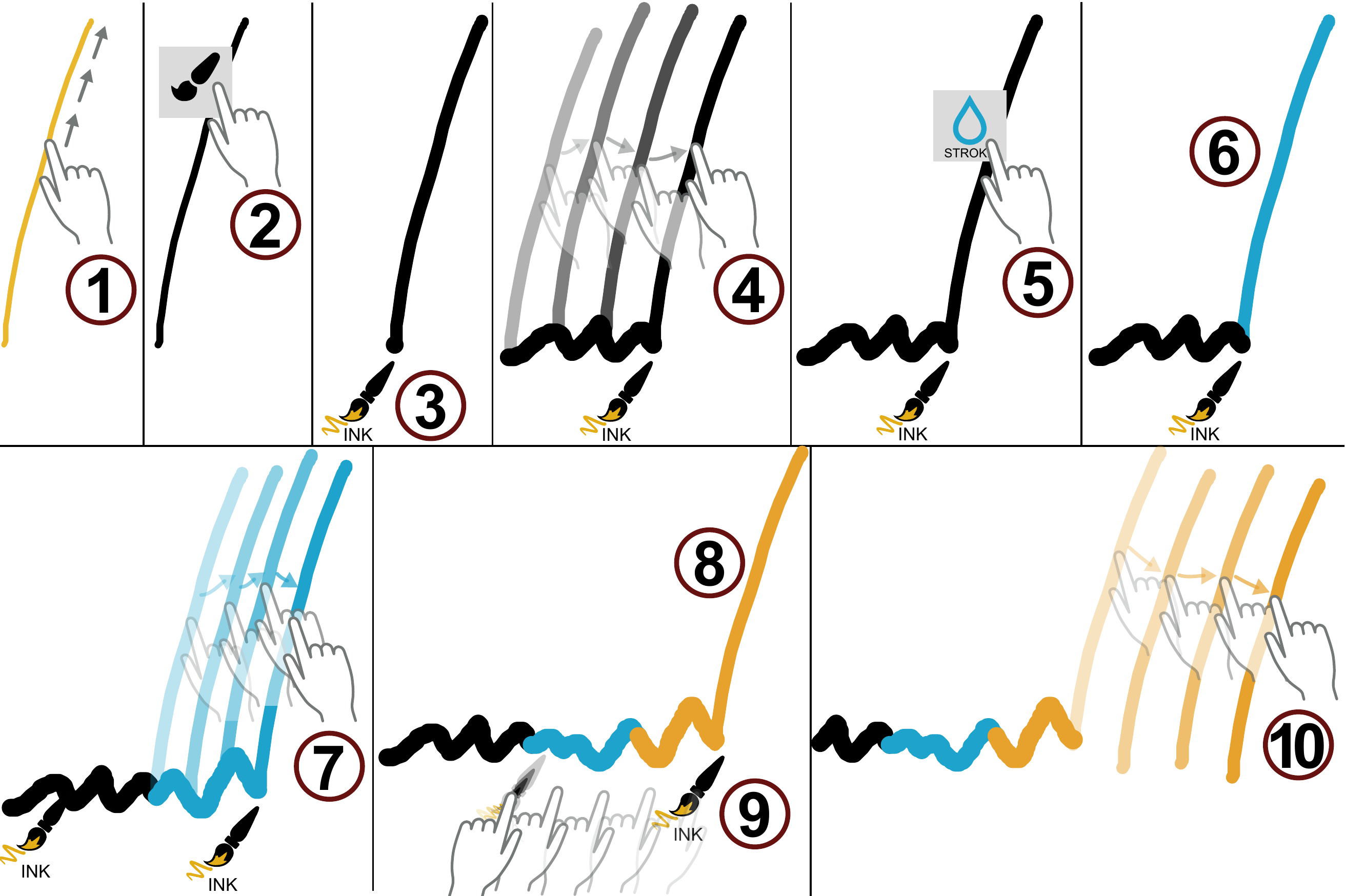}
		\caption{
			Constructing a \quotes{pen} tool from a user-drawn stroke (1) by attaching an \propCaption{ink} activator (2). The pen's (3, 4) ink is configured via \propCaption{color} modifiers (5--9).
		}
		\label{fig:constructingAPen}
	\end{center}
	\vspace{-2.5em}
\end{figure}

\begin{figure*}[b!]
	\begin{center}
		\includegraphics[width=\textwidth]{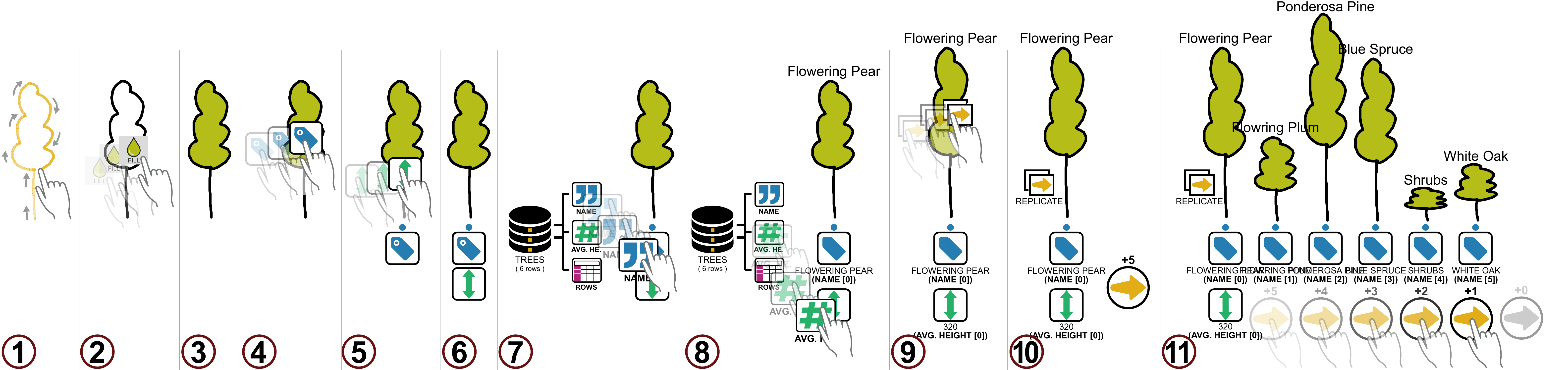}
		\caption{
			Constructing a bar chart using marks in the shape of trees. 
			After drawing the outline of a tree (1), dropping the transient \propCaption{fill} color modifier (2) changes the fill color of the mark (3). 
			Adding a \propCaption{label} (4) and a \propCaption{height} (5) modifier attach persistent representations of these modifiers underneath the mark (6), without any effect on the mark itself.
			Mapping the \propCaption{name} attribute to the \propCaption{label} modifier (7) adds the name of the first tree in the dataset at the top of the mark (8).
			Mapping the \propCaption{avg.height} attribute to the \propCaption{height} modifier (8) sets the height of the mark to the average height of the first tree in the dataset (9).
			Adding a \propCaption{replicate} activator to the tree and creates an interactive handle (10) that can be dragged right to create new visual mark that replicate the mappings of the original mark.
		}
		\label{fig:treechart-mark-bar}
	\end{center}
	\vspace{-2em}
\end{figure*}

\begin{figure}[t!]
	\begin{center}
		\includegraphics[width=\columnwidth]{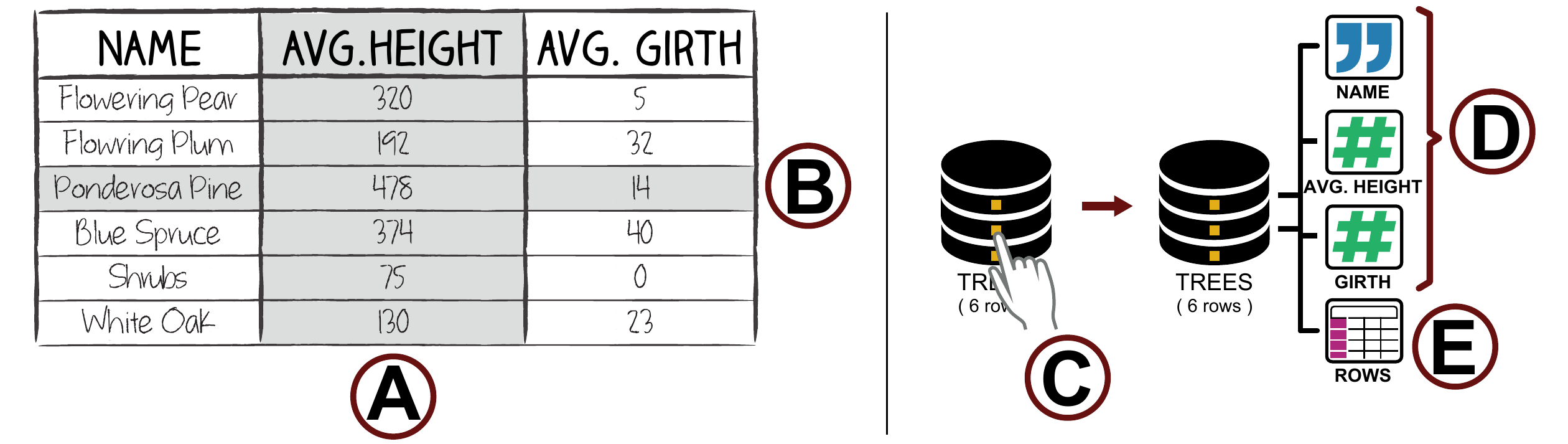}
		\caption{
			A tabular dataset about trees, deconstructed into attributes (A) and records (B). 
			This dataset can have a compressed representation (C), or an expanded representation showing its attributes (D) and records (E).
		}
		\label{fig:datasetVisualRepresentation}
	\end{center}
	\vspace{-2.75em}
\end{figure}

\subsection{An Illustrative Example}

To illustrate how we can use activators and modifiers to turn a passive object into an active tool, we describe the construction of a \quotes{pen} (Fig.~\ref{fig:constructingAPen}). In this case, we do not use any data. Instead, we simply illustrate the interaction aspects of our constructive approach through construction of a drawing tool. Here modifiers and activators have graphical representations that suggest their type and function and are added to objects via drag-and-drop gestures.\\ 

\noindent We first add to a user-drawn line (1) an \prop{ink} activator (2), which in this scenario carries a \prop{draw} function. The added activator appears as a directly manipulable graphical element that is attached to the activated stroke (3). The \prop{ink} activator of our example gives objects the ability to draw on the canvas when moved, converting the activated line into a \quotes{pen}. In this example, a pen generates scribbles on the canvas with the same width and color of the activated line (4). We then use a \prop{stroke} modifier (5) to change the activated line's stroke color to blue (6). Consequently, new movements of the pen produce blue scribbles (7). Subsequent modifications of the object's stroke color will also change the color of the pen's ink (e.g., to orange--8).
Dragging the brush icon out of our pen tool deactivates it (9). This step does not affect the line's current visual appearance or the scribbles previously drawn with it. Once deactivated, the stroke no longer draws when moved (10).

\begin{figure*}[t!]
	\begin{center}
		\includegraphics[width=\textwidth]{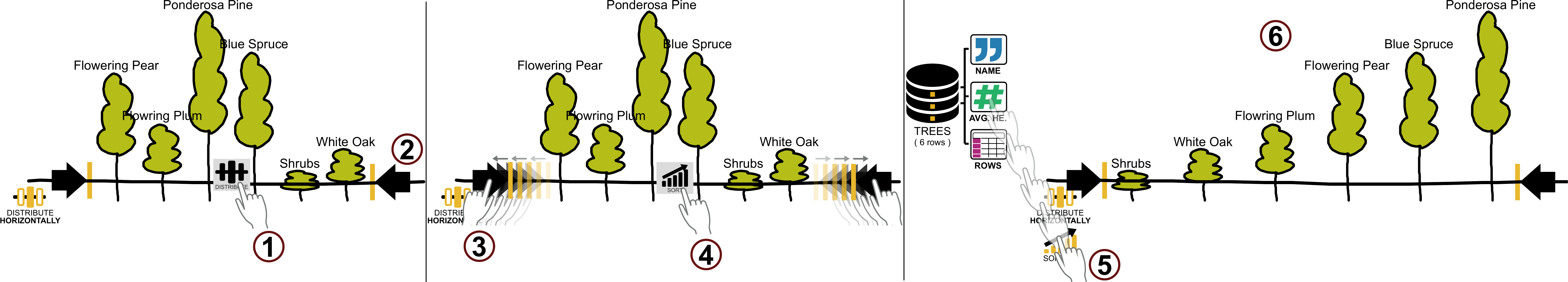}		
		
		\caption{
			Distributing and sorting the marks of a chart. 
			Drawing a horizontal stroke on top of the trees and adding a \propCaption{distribute} activator to that stroke (1) turns the stroke into a tool.
			This tool shows two handles (2) that can be dragged outward (3) to distribute and increase the horizontal space between marks.
			Adding a \prop{sort} activator (4) to the tool and mapping the \propCaption{avg.height} data attribute to it (5) sorts the marks based on their \propCaption{avg.height} value, in ascending order (6).
		}
		\label{fig:treechart-distribute-sort}
	\end{center}
	\vspace{-2em}
\end{figure*}

%% file: sections/05.SystemDescription.tex
\section{\app}
\label{section:VisAssembler}

We now explain how objects, modifiers, activators and tools integrate in \app~to support visualization creation.

\subsection{The Interactive Environment}

In \app, visualizations are constructed within a canvas where people draw and move strokes via touch, pen or mouse. These strokes are the \textit{objects} of our domain. We use \textit{modifiers} to change objects' visual appearance. \textit{Activators} add functionality to objects, which enables the construction of tools to operate on other objects (e.g., a tool to spatially arrange a set of rectangles). As encouraged by the idea of Instrumental Interaction~\cite{BeaudouinLafon:2000wj}, \app~pays attention to visibility and enables direct manipulation~\cite{Shneiderman1983} of its interaction elements. Thus, both activators and modifiers are available as icons in a palette and are attached to (or removed from) objects via drag-and-drop---as in the illustrative example. When added to an object, activators are represented graphically, with an icon that suggests their function~\cite{norman:1988:psychology_everyday_things}. The same applies for persistent modifiers. Transient modifiers produce an immediate change on an object's visual appearance.

\subsection{Working with Data}

\app~currently supports tabular datasets. 
We deconstruct tabular datasets into their structural components: data dimensions (i.e. columns) and records (i.e. rows). 
A data dimension is a collection of all the values of a given column and is named after the column (Fig.~\ref{fig:datasetVisualRepresentation}.A). 
A record is a set of attribute-value pairs (e.g., the record \prop{\{name: `Ponderosa Pine', avg.height: 478, avg.girth: 14\}} shown in Fig.~\ref{fig:datasetVisualRepresentation}.B) In \app, a dataset is represented on the canvas either in compressed (Fig.~\ref{fig:datasetVisualRepresentation}.C) or expanded form (Fig.~\ref{fig:datasetVisualRepresentation}.D). 
When expanded, the dataset provides access to representations of its data dimensions (Fig.~\ref{fig:datasetVisualRepresentation}.D) and records (Fig.~\ref{fig:datasetVisualRepresentation}.E). These representations are draggable objects whose icon and color suggest the underlying data type: blue half-quotes for categorical attributes and green pound sign for quantitative ones. \app~organizes the dataset's components in an ordered sequence. 
For example, the first value of the \prop{Name} dimension is \prop{Name[0]} and the first record of the dataset is \prop{Rows[0]}. 
This order is relevant for some activators such as the \prop{tuple} and \prop{replicate} activators, as explained later.

\section{Constructing a Bar Chart}

We now use a running example of constructing a sorted bar chart made of tree shapes to show the use of our interaction elements.

\noindent A bar chart typically consists of two or more bars which: 
\begin{enumerate*}
	\item have a height that represents an aspect of the data,
	\item are aligned to a common base,
	\item are optionally spaced for readability, and
	\item ideally, are labeled according to the data they present.
\end{enumerate*}\\

\noindent
\textbf{Step 1: Creating a visual mark with visual properties.} 
Marks can be created from any object drawn on the canvas. 
These objects have inherent attributes such as \prop{width}, \prop{height}, and \prop{stroke} and \prop{fill} colors.
Fig.~\ref{fig:treechart-mark-bar} (1--3) shows how to construct a visual mark from a hand-drawn stroke using the \prop{fill} transient modifier to set the fill color property of the tree to green. Other transient modifiers such as \prop{stroke} color and \prop{shape} beautification can be applied similarly.\\

\noindent
\textbf{Step 2: Mapping data dimensions to visual properties.} 
We now show the label associated to the data record our tree mark represents, as well as map its height value to the mark's height attribute.
Fig.~\ref{fig:treechart-mark-bar} (4--9) shows the steps involved in establishing these data mappings.
We first add two persistent modifiers to the mark: \prop{label} (4) and \prop{height} (5). 
Because no data is mapped to either modifier, there is no effect on the mark itself but icons representing these modifiers are attached underneath the mark (5, 6). These icons can be used to map different data dimensions to visual properties.
To establish a data mapping, we drag a data dimension from the dataset and drop it onto the icon of a persistent modifier. When mapping the \prop{Name} attribute to the \prop{label} modifier of the mark (7), the name of the first record in the dataset (\prop{Name[0]}) is attached to the tree (8).
We then map the \prop{Avg.Height} data dimension to the \prop{height} modifier (8). This changes the height of the mark using the value of the first record in the dataset (\prop{Avg.Height[0]}).\\

\noindent
\textbf{Step 3: Replicating the defined mappings.} 
At this stage, we have created a single mark that represents the first record of our dataset. 
Fig.~\ref{fig:treechart-mark-bar} (9--11) shows how we use a \prop{replicate} activator to avoid having to manually define mappings for each record of the dataset.
We first add the \prop{replicate} activator to the mark (9). This adds a handler to the right of the mark's visual properties (10, yellow encircled arrow). This step turns the newly activated object into an interactive tool that can replicate itself.
A numeric value on top of the yellow handler indicates the number of additional data records that can be replicated. 
We then drag the handler from left to right to create new marks (11). The more we drag, the more visual marks are created. 
Each newly created mark replicates the mappings of the original one using subsequent data records (e.g., the first replicated mark uses the values \prop{Name[1]} and \prop{Avg.Height[1]}). 
As new marks are replicated the number of available records to replicate shown on top of the handler decreases. 
When this value is zero the handler is grayed out and cannot be dragged  further.\\

\noindent
\textbf{Step 4: Distributing and sorting the trees.} 
\label{section:distributingAndSorting} 
Now that we have constructed a bar chart, we want to rearrange the trees to evenly add some space between them, and to sort them in ascending order.
We achieve these actions by creating another interactive tool, as shown in Fig.~\ref{fig:treechart-distribute-sort}.
We draw a horizontal stroke on top of the trees and add a \prop{distribute} activator to it (1).
Upon adding the \prop{distribute} activator, two interactive handles appear at the start and end points of the space spanned by the marks intersected by the stroke (2). 
Activating the stroke turns it into a distributing \textit{tool}. Moving its handles (3) increases (or decreases) the horizontal space between the trees.
The last step is to sort the trees in ascending order.
We add a \prop{sort} activator (4) to the horizontal stroke and drag the \prop{Avg.Height} data dimension of the dataset onto it (5). 
This sorts the trees in ascending order by default (6). Tapping the \prop{sort} activator's icon changes the sorting direction (to descending) if needed.

%% file: sections/09.Discussion.tex
\section{Discussion}

We chose to explore a variation of a digital constructive approach to creating visualizations. A constructive approach requires components from which the visualization can be built. Previous constructive approaches~\cite{Huron.2014a, Mendez2016} have decomposed the data into individual data entities which can be represented as tokens. Construction can then proceed with these tokens. While this has shown to empower people and promote data and visualization understanding~\cite{huron.2014b, Mendez2017, Wun2016}, people also object to the tedium of moving individual tokens~\cite{Wun2016}.\\

\noindent In \app~we consider leveraging the tabular data structure in our deconstruction. This approach to deconstruction, allows the user to make use of this structure through our replicator activator to  pull related data entities unto the canvas. This helps alleviate some of tedium of placing the single data items, and points to a possible direction for combining the advantages of constructive visualization with a more scalable interactive approach.

\subsection{Flexibility and Sequencing}

One of the key points of construction is the potential to achieve similar outcomes by combining atomic elements in multiple ways. In contrast to tangible tools, this is particularly true for digital construction, as software-supported processes can be notoriously mutable. Our ongoing observations show that people expect a more relaxed workflow regarding the order of operations they executed in \app. This is in line with studies that have shown that, when the freedom is available,
humans do not follow particular sequences during design but rather
opt for personal variations that change with the problem at hand [19]. Developing \app, however, made evident that full flexibility can be hard to achieve in a software solution. It appears that some order is required to achieve a construction. In \app we make use of the order used to store the tabular data.

\subsection{Granularity, Scalability and Agency}

\app~operates on data dimensions—rather than individual values like other constructive tools. In combination with our replication strategy, working at this coarser granularity level provides some scalability. We also circumvent scalability issues by enabling people to create tools to operate on the tokens they use. One can construct, for example, a \quotes{sorting tool} to sort the marks of a visualization according to a specific data dimension. This is a unique aspect of \app that places more responsibility (hence, agency) in the hands of the designer. It also differs from most conventional visualization tools in that although certain operations can be speeded up, they are not fully delegated to the tool but rather constructed by the designer.

%% file: sections/10.Conclusion.tex
\section{Conclusion}

\app~provides a new variation on constructive visualization that supports the creation of visualizations based on four reusable interaction elements---objects, modifiers, activators and tools. In creating a visualization with \app~visual marks are constructed by attaching modifiers to hand-drawn canvas \textit{objects}. Persistent \textit{modifiers} provide mechanisms for establishing data bindings, achieving data-driven modifications of an object's visual appearance. \textit{Activators} can also be added to the visual marks to enable the construction of \textit{tools} that can operate on the visualization's components. In combination, these four interaction elements enable the creation of visualizations through a constructive authoring process.

%
%

%% file: sections/11.Acknowledgements.tex
\acknowledgments{
This project was funded in part by the National Sciences and Engineering Research Council of Canada, Alberta Innovates Technology Futures, SMART Technologies, and the European Union's Horizon 2020 research and innovation programme under the Marie Sklodowska-Curie grant agreement No. 753816.
}